\documentclass[aps,floats,prl,showpacs,twocolumn ]{revtex4}

\usepackage{amsfonts,amsmath} \usepackage{bm} \usepackage{dcolumn}
\usepackage{epsfig} \usepackage{latexsym}

\begin{document}

\title{Light localization signatures in backscattering from periodic disordered media}

\author{Chushun Tian}

\affiliation{Institut f{\"u}r Theoretische Physik, Universit{\"a}t
zu K{\"o}ln, K{\"o}ln, 50937, Germany}

\date{\today}
\begin{abstract}
 {\rm The backscattering line shape is analytically predicted for
 thick disordered medium films where, remarkably, the medium configuration is periodic along the direction
 perpendicular to the incident light. A blunt triangular peak is found to emerge on the
 sharp top. The phenomenon roots in the coexistence of quasi-$1$D localization
 and
 $2$D extended states.
 }
\end{abstract}


\maketitle

{\it Introduction.} Since the mid-eighties \cite{Golubentsev}
the coherent backscattering (CBS) has been one of the pilots of
studies of Anderson localization of light. Indeed, manifestations of
weak localization (WL) in the CBS line shape have been well
documented for various disordered dielectric media
\cite{Niuwenhuizen},
while how strong localization (SL) affects CBS has been a long term
fascinating subject \cite{Berkovits87}. The last decade has
witnessed spectacular progress on CBS near
\cite{Lagendijk97,Lagendijk00} or far below \cite{Zhang02} the
localization transition, which undoubtedly is an intellectual
challenge both experimentally and theoretically. Indeed, to prepare
strong scattering media and to extract localization from medium
absorption are highly restrictive \cite{Lagendijk97}, while the
failure of perturbation theory \cite{Efetov97}--crucially mapping
the pictorial reciprocal paths into (diagrammatical) one-loop
approximation \cite{Golubentsev,Niuwenhuizen}--enforces the
invention of a nonperturbative theory to allow a microscopic
analysis. Despite of these difficulties a common belief is that SL
is responsible for rounding the CBS sharp top
\cite{Berkovits87,Lagendijk97,Lagendijk00,Zhang02}.
\begin{figure}
\begin{center}
\leavevmode \epsfxsize=8cm \epsfbox{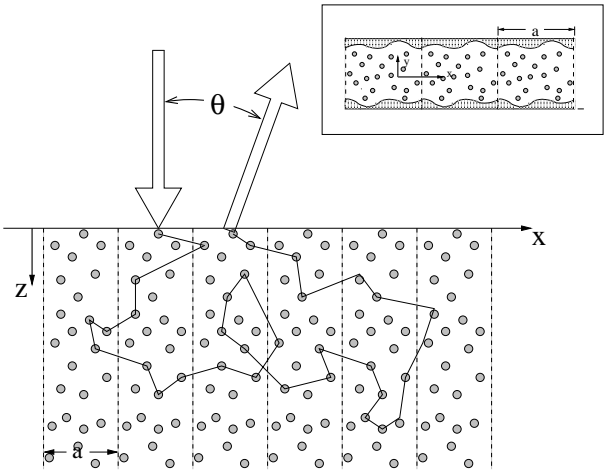}
\end{center}
\caption{Light backscattering from a periodic thick disordered
medium film. Inset: the film section.}
  \label{fig1}
\end{figure}

In studies of light localization much attention has been paid to
fully disordered media. There have also been increasing interests on
other medium structure such as disordered photonic crystals
\cite{Vos00} where the Bloch symmetry is slightly destroyed by
impurities, and systems with perfect Bloch symmetry \cite{Tian05}
where WL is analytically found. Most interestingly, the recent
invention of so-called planar random laser introduces a novel medium
structure \cite{Xu06}, which consists of a random gain layer
sandwiched by two mirrors. It was then conjectured that SL in the
layer plane might be responsible for the laser emission \cite{Xu06}.
To prove it is yet a nontrivial task which may be traced back to the
striking feature of the partially disordered structure. Indeed,
(from the geometrical optics view) two perfect reflection mirrors
map the medium to an extended periodic one but fully disordered
inside the primitive cell (apart from the mirror symmetry
). This immediately raises many important questions. For examples,
{\it does Anderson localization exist in such structure? If so, can
it be probed by CBS measurements?}

Unfortunately, the interplay between Anderson localization and the
periodicity is a difficult issue. In general the common wisdom
regarding localization
may be drastically modified and very little has been known, among
which are: a new scale essential to WL
\cite{Tian05} may appear, and whether the constructive interference
between reciprocal paths encompasses WL depends on periodic medium
configuration \cite{Tian05,Altshuler93}.

In this letter these two problems will be investigated for a
simplified model--thick disordered medium film with a dielectric
function {\it periodic} in the $x$-direction (Fig.~\ref{fig1}). The
lattice constant is $a$ and the primitive cell
consists of randomly positioned point-like scatterers
filling the half space $z>0$\,. The film section is uniformly
illuminated by a beam of stationary unpolarized light (with the
wavelength $\lambda$ and the frequency $\Omega$) perpendicular to
it.
Microscopic analysis is provided for the angular resolution of light
intensity near the inverse incident direction for sufficiently large
times.

\begin{figure}
\begin{center}
\leavevmode \epsfxsize=8cm \epsfbox{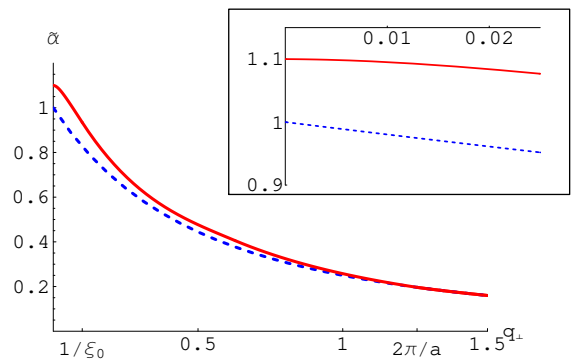}
\end{center}
\caption{The predicted (solid) versus traditional (dotted) line
shape (symmetric with respect to $q_\perp =0$). The parameters are
$a=5\, l,\, \xi_0 = 10\, l$ (setting $l=1$). Inset: the blunt
triangular peak at $|q_\perp|\ll \xi_0^{-1} = 0.1$\,.}
  \label{fig2}
\end{figure}

Qualitatively, as shown in (Fig.~\ref{fig2}), the traditional line
shape \cite{Golubentsev,Niuwenhuizen} develops at $|q_\perp|\geq
2\pi/a$ and is sharpened at $\xi_0^{-1} \lesssim |q_\perp|<
2\pi/a$\,, eventually a blunt triangular peak emerges at $|q_\perp|
\ll \xi_0^{-1}$\,. Here $q_\perp =(2\pi/\lambda)\sin\theta$ and
$\xi_0=\pi a \nu D$
with $\nu$ the photon density of states at $\Omega$ and $D=l/3$ the
bare diffusion constant ($l$ the transport mean free path and the
velocity $c=1$). Quantitatively, analytical predictions are made for
$\lambda \ll l \ll a \ll l \, e^{l/\lambda}$\,, where the last
inequality ensures photon states to be far from $2$D SL
\cite{Efetov97}. In particular, the line shape
$\tilde{\alpha}(\theta) $ is singular at $|q_\perp| =0\,, 2\pi/a$\,,
around which
\begin{eqnarray}
\tilde{\alpha}(\theta) = \Bigg\{\begin{array}{c}
  (1+l |q_\perp|)^{-2} \equiv I_0(\theta)\,,\,
  \qquad
  |q_\perp|>\frac{2\pi}{a} \,; \\
\{1+\gamma(\theta)\}\, I_0(\theta)\,, \quad
0<\frac{2\pi}{a} - |q_\perp| \ll \frac{\pi}{a}
\,; \\
\left(1+l/\xi_0\right)-a/(2\pi\xi_0)\, l|q_\perp|  \,, \quad
|q_\perp| \ll \xi_0^{-1} \,, \\
\end{array}
\label{result}
\end{eqnarray}
while a smooth interpolation between the last two lines is expected.
Here the enhancement factor $\gamma (\theta) =
(3l/2\xi_0) \, [1-a|q_\perp|/(2\pi)]$\,.

{\it Qualitative picture.} The well known Bloch theorem allows us to
reduce the photon motion into an effective one
within a primitive cell dictated by the Bloch wave number $k_\rho
$\,. The enhanced backscattering finds its origin
in the constructive interference--described by low-energy
hydrodynamic modes--between two counter-propagating photons
(so-called cooperon) each of which carries a Bloch wave number
$k_\rho\,, k_\rho'$\,, respectively. $k_\rho+k_\rho'$ plays the role
of ``Aharonov-Bohm flux''. The gauge invariance then requires that
the transverse ($x$-direction) hydrodynamic wave number $2\pi N/a$
with $N$ an integer satisfies $q_\perp=2\pi N/a-(k_\rho+k_\rho')$\,.
Hydrodynamic modes with $N\neq 0$ ($N=0$) describe $2$D (quasi-$1$D)
motion.

As $k_\rho\,, k_\rho'\in [-\pi/a, \pi/a)$ there are two
contributions which correspond to two successive $N$ responsible for
the line shape. For $|q_\perp|>2\pi/a$ both $N\neq 0$\,. Therefore,
these two hydrodynamic modes are inhomogeneous in the transverse
direction, and extended in the longitudinal ($z$-) direction because
of $a \ll l \, e^{l/\lambda}$\,. As usual the diffusive $2$D motion
then leads to the traditional line shape.

For $|q_\perp|<2\pi/a$ the line shape is contributed by both $2$D
diffusive and quasi-$1$D motion since the two successive integers
are now $1$ (or $-1$) and $0$\,. The former (or latter) occupies a
portion of $a |q_\perp| /2\pi$ (or $1 - a |q_\perp| /2\pi$). In the
quasi-$1$D geometry there are two scales: $l$ and $\xi_0$\,. An
incident flux decays over the scale $\sim l$ then diffuses inside
the medium and eventually exit at $z\sim l$\,. The penetration
length is $\sim |q_\perp|^{-1}\gg l$\,.

\begin{figure}
\begin{center}
\leavevmode \epsfxsize=8cm \epsfbox{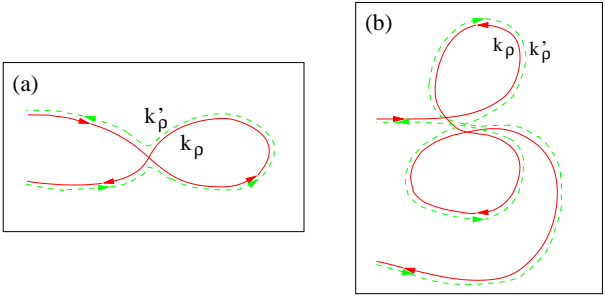}
\end{center}
\caption{One-loop (a) and typical two-loop (b) interference picture
underlying quasi-$1$D weak localization.}
  \label{fig3}
\end{figure}

For $|q_\perp|$ closed to $2\pi/a$ photons penetrate into the medium
of a distance $
\ll \xi_0$ via quasi-$1$D diffusive motion. Upon penetration they
may self-intersect then propagate around the formed loop along the
same direction--so-called diffuson (Fig.~\ref{fig3} (a)). The
probability is larger in quasi-$1$D than in $2$D. As a result two
initially closed but counter-propagating photons have a larger
probability to be brought back to their starting point and the
backscattered light intensity is thereby enhanced. At $z\sim l$ the
probability of forming such a loop is $\sim l/D$\,. Consequently,
the inverse diffusivity at $z\sim l$ increases with an amount of
$l/(a\nu D) \sim l/\xi_0$\,. Noticing that the backscattered light
intensity is proportional to the inverse diffusivity $D^{-1}$
\cite{Golubentsev} and taking into account the weight of quasi-$1$D
motion, we find that the traditional line shape is magnified by a
factor of $1+\gamma(\theta)$\,.

Notice that the quasi-$1$D cooperon and diffuson have different
masses: $D(k_\rho+k_\rho')^2=Dq_\perp^2$ and
$D(k_\rho-k_\rho')^2$\,, respectively. As $|q_\perp|$ decreases the
diffuson tends to acquire a larger massive and be damped.
Consequently, all the constructive interference involving
diffuson-cooperon coupling (e.g., Fig.~\ref{fig3} (a)) tends to be
suppressed. Opposed to this the cooperon becomes less massive.
Consequently higher order loop-wise interference paths involving
solely cooperon-cooperon coupling (e.g., Fig.~\ref{fig3} (b))
accumulate and eventually dominate the backscattered light intensity
at $|q_\perp| \lesssim \xi_0^{-1}$\,, where photons penetrate deeply
into the medium forming SL in the bulk.

In this region from one-parameter scaling hypothesis we expect that
the diffusion coefficient exponentially decays from the interface,
i.e., $-\ln D(z)\propto z/\xi_0$\,. Therefore, the average inverse
diffusivity of the boundary layer increases also by an amount of
$\sim l/\xi_0$\,. Thus for $|q_\perp| \ll \xi_0^{-1}$ the quasi-$1$D
motion contributes to the line shape $(1+l/\xi_0)[1 - a |q_\perp|
/(2\pi)]$\,, together with the portion contributed by the $2$D
extended motion: $I_0(\theta) [a |q_\perp| /(2\pi)]\approx a
|q_\perp| /(2\pi)$\,, and leads to a blunt triangular peak.

{\it General formalism.} We then outline the proof
\cite{notecondition}. Let us start from the retarded (advanced)
Green's function: $ G^{R,A}_{\Omega^2}({\bf R},{\bf
 R}')=\langle {\bf R} |
 \{\nabla^2+ \Omega^2\left[1+\epsilon({\bf R})\right]\pm i0^+\}^{-1}
  |{\bf R}' \rangle$ describing the propagation of the electric
  field.
The fluctuating dielectric field
$\epsilon({\bf R})$ has zero mean and vanishes for $z<0$\,, while
for $z>0$ is periodic in $x$
and satisfies $\Omega^4 \left\langle \epsilon({\bf R})\epsilon({\bf
R}')\right\rangle = \Delta \, \delta(y-y')\,\delta(z-z')\,
\sum_{N\in {\mathbb{Z}}}\, \delta (\rho-\rho'-Na)$\,, where
$(\exists m\in \mathbb{Z})\, \rho = x-ma \in [0,a)$ stands for the
relative $x$-coordinate in the primitive cell. The scattering is
elastic with $l =4\pi/\Delta$\,. Using
the effective medium approximation the incident field is ${\bar
E}({\bf R}) =E_0 e^{i\Omega{\hat {\bf z}}\cdot {\bf R}-z/2l}$
\cite{Stephen86}.
Alternatively, one may introduce the
Green's functions: ${\cal G}^{R,A}_{\Omega^2}({\bf r},{\bf
r}';k_\rho) = \langle{\bf r}|(\Omega^2 - {\hat H}(k_\rho) \pm
i0^+)^{-1}|{\bf r}'\rangle$ for the effective motion within the
primitive cell with ${\hat
H}(k_\rho)=-[(\partial_\rho+ik_\rho)^2+\partial_y^2+\partial_z^2]-\Omega^2
\, \epsilon ({\bf r})$ (${\bf r} \equiv (\rho,y,z)$).
The two sets of Green's functions are related through
\begin{equation}
G^{R,A}_{\Omega^2}({\bf R},{\bf R}') = \sum_{k_\rho} \,
e^{ik_\rho(x-x')} {\cal G}^{R,A}_{\Omega^2}({\bf r},{\bf r}';k_\rho)
\,. \label{Greenrelation}
\end{equation}

The albedo characterizing the radiation intensity in the direction
${\bf s}= (\sin\theta\,, 0\,,-\cos\theta)$ \cite{Stephen86}
generally depends on the time $t$\,. With the single scattering
event ignored the albedo at $t\rightarrow \infty$\,, denoted as
$\alpha(\theta)$\,, can be shown to be
\begin{eqnarray}
\alpha(\theta) & = & \int\!\!\!\!\int d{\bf R}_1d{\bf R}_2 \,
e^{-\frac{z_1+z_2}{l} } \Large[\overline{G^R_{\Omega^2} ({\bf
R}_1,{\bf
 R}_2) \, G^A_{\Omega^2} ({\bf R}_2,{\bf R}_1)} \nonumber\\
 & + & e^{i\Omega {\bf s}\cdot ({\bf
R}_2-{\bf R}_1)}\, \overline{G^R_{\Omega^2} ({\bf R}_1,{\bf R}_2) \,
G^A_{\Omega^2} ({\bf R}_1,{\bf R}_2)}\Large] \label{J2}
\end{eqnarray}
in the unit of $\Delta^2 E_0^2/16 \pi$ with $\overline {(\cdots)}$
the average over the fluctuating dielectric field $\epsilon({\bf
R})$\,. The first (second) term gives the background intensity
$\alpha_0$ (line shape $\tilde{\alpha}(\theta)$). Notice that the
$y$-dependence of Green's functions is irrelevant and will be
ignored from now on.
Substituting Eq.~(\ref{Greenrelation}) into Eq.~(\ref{J2}) gives (up
to an irrelevant overall normalization factor)
\begin{eqnarray}
\!\!\!\!\!\!\alpha_0 &  =  & \!\! \int\!\!\!\!\int_0^\infty dzdz'
e^{-\frac{z+z'}{l}} \sum_{k_\rho} {\cal Y}^{\rm {D}}_0
(z,z';k_\rho,k_\rho), \label{background}
\\
\!\!\!\!\!\!\tilde{\alpha}(\theta) &  =  & \!\!\int\!\!\!\!\int
_0^\infty dzdz' e^{-\frac{z+z'}{l}}\sum_{N}
\sum_{k_\rho,k_\rho'}\!'^N \,{\cal Y}^{\rm {C}}_N
(z,z';k_\rho,k_\rho'), \label{albedo1}
\end{eqnarray}
where we have quantified the propagators: diffuson and cooperon
introduced above to be ${\cal Y}^{\rm {D}} ({\bf r},{\bf
r}';k_\rho,k_\rho') \equiv \overline{{\cal G}^R_{\Omega^2} ({\bf
r},{\bf r}';k_\rho) \, {\cal G}^A_{\Omega^2} ({\bf r}',{\bf
r};k_\rho')}$ and ${\cal Y}^{\rm {C}} ({\bf r},{\bf
r}';k_\rho,k_\rho') \equiv \overline{{\cal G}^R_{\Omega^2} ({\bf
r},{\bf r}';k_\rho) \, {\cal G}^A_{\Omega^2} ({\bf r},{\bf
r}';k_\rho')}$\,,
and introduced their Fourier transformations, i.e., ${\cal Y}^{\rm
{D,C}} ({\bf r},{\bf r}';k_\rho,k_\rho')=a^{-1}\sum_{N\in
\mathbb{Z}}\, e^{i(\rho-\rho')2\pi N/a }\, {\cal Y}^{\rm {D,C}}_N
(z,z';k_\rho,k_\rho')$\,. The partial summation
$\sum_{k_\rho,k_\rho'}^{'N} \equiv \sum_{k_\rho,k_\rho'}
\delta_{k_\rho+k_\rho' \,, 2\pi N/a - q_\perp}$\,.

One may proceed to
sum up over all the (maximally crossing) ladder diagrams (e.g.,
Refs.~\cite{Efetov97,Stephen86}) producing a bare diffuson
(cooperon) ${\cal Y}^{\rm {D(C)}}$\,. Nevertheless such a
diagrammatic expansion fails in the nonperturbative
analysis--intrinsic to localization. Instead, for interactionless
systems such as photons to fulfill this task the supersymmetric
method--to which we switch below--turns out to be perfectly suitable
\cite{Efetov97}.
Conceptually, the introduced
slow varying $Q$-field interprets the bare diffuson and cooperon as
Goldstone modes of the spontaneous supersymmetry breaking, and
encapsulates their mutual interactions underlying localization
through the nonlinear constraint: $Q^2=1$\,. Technically, it can be
shown
 \begin{eqnarray}
\!\!{\cal Y}^{\rm {D,C}} ({\bf r},{\bf r}';k_\rho,k_\rho') =
\frac{1}{2^7}\langle{\rm str} [k\Lambda^+ \tau^- Q({\bf r})
\Lambda^- \tau^\mp kQ({\bf r}')] \rangle
\label{DC}
\end{eqnarray}
following standard derivations \cite{Efetov97}, where $Q({\bf r}) =
T^{-1}({\bf r}) \Lambda T({\bf r})$ is
a matrix field
with full orthogonal symmetry, i.e., $T \in U(2,2/4)/U(2/2)\times
U(2/2)$\,. The average is defined as $\langle\dots\rangle \equiv
\int \, D[Q] (\cdots) e^{-F[Q]}$ with $Q|_{z=-z_0}=\Lambda,
Q|_{\rho=0}=Q|_{\rho=a}$\,, where the action $F[Q]=\frac{\pi\nu
D}{8}\int_0^\infty dz\int_0^a d\rho\, {\rm str} [(\partial_z Q)^2 +
(D_\rho Q)^2]$
with
$D_\rho \equiv \partial_\rho + i [
{\bar{\bar k}} \tau_3,\cdot ]$\,. ${\rm str}$ is the supertrace and
all the matrices follow the definitions of
Ref.~[\onlinecite{Efetov97}]\,. In addition,
$\Lambda^\pm = (1\pm \Lambda)/2$\,, $\tau^\pm = (1\pm \tau_3)/2$\,,
and the matrix $\overline{\overline{k}} = {\rm diag}(k_\rho\,,
k_{\rho}')$ is diagonal in the retarded-advanced sector.
The boundary condition $Q(-z_0)=\Lambda$ accounts for the fact that
the albedo is contributed by optical paths not crossing the trapping
plane located at $z=-z_0\approx -0.7 \, l\approx 0$
\cite{Golubentsev}, which coincides with the medium boundary.

Eqs.~(\ref{background})-(\ref{DC}) constitute the general formalism
of calculating the albedo. By making an appropriate global rotation
for the $Q$-field reflecting the symmetry: ${\cal Y}^{\rm {D}} ({\bf
r},{\bf r}';k_\rho,k_\rho')={\cal Y}^{\rm {C}} ({\bf r},{\bf
r}';k_\rho,-k_\rho')$\,, it can be shown that, similar to the fully
disordered medium with conserved reciprocity \cite{Golubentsev}, the
background intensity $\alpha_0 = \tilde{\alpha}(0)$\,. We then turn
to analyze the line shape.

{\it Line shape.} Three regions: (i) $|q_\perp| \geq 2\pi/a$\,, (ii)
$\xi_0^{-1}\lesssim |q_\perp| < 2\pi/a$ and (iii) $|q_\perp|
\lesssim \xi_0^{-1}$ will be studied separately. It can be shown
that in (i) and (ii) $Q({\bf r})$ mildly fluctuates around
$\Lambda$\,. A perturbation theory near it generates leading
diffusive motion--described by the bare diffuson (cooperon)--and
loop-wise interference (Fig.~\ref{fig3}) encompassing WL. Opposed to
this, in (iii) though locked at $\Lambda$ at
the boundary $Q({\bf r})$
may strongly fluctuate along the $z$-direction driving photons into
SL states, while remains homogeneous in the $\rho$-direction\,.
Consequently the perturbation theory breaks down.

In Eq.~(\ref{albedo1}) the summation over $N$ picks up two terms
with successive integers $N=N',N'+1$\,. In (i) they both do not
vanish corresponding to $2$D low-energy extended motion. The leading
order perturbation then gives
\begin{eqnarray}
\!\!\!\! \left\{\!-\partial_z D \partial_z + D \!\left[\frac{2\pi
N}{a} + (k_\rho\mp k_\rho')\right]^2 \! \right\} {\cal Y}^{\rm
{D,C}}_N
(z,z';k_\rho,k_\rho') \nonumber\\
=\frac{1}{\pi\nu}\, \delta(z-z')\,,\qquad {\cal Y}^{\rm
{D,C}}_N|_{z=0 }=0 \,.\quad \label{bareproagator}
\end{eqnarray}
Substituting its solution into Eq.~(\ref{albedo1}) recovers
the first line of Eq.~(\ref{result}) \cite{notefactor,noteWL}.
In (ii) and (iii) $|N|=0\,, 1 $\,. Taking into account the weight of
$2$D motion, i.e., $a|q_\perp|/(2\pi)$ the term with $|N|=1$ is
found to be $ \frac{a|q_\perp|}{2\pi}\, I_0(\theta)$\,.
The term with $N=0$ arises from the quasi--$1$D motion suffering
from localization effects, and below will be calculated separately
for (ii) and (iii) with the simplified action $F[Q]=\frac{\pi a\nu
D}{8}\int_0^\infty dz \, {\rm str} \{(\partial_z Q)^2 +
[i\bar{\bar{k}}\tau_3, Q]^2 \}$ (since $Q$ does not depend on
$\rho$).

For (ii) one-loop expansion show that Eq.~(\ref{bareproagator})
($N=0$) still holds reflecting the flux conservation. However, the
diffusion coefficient acquires the position-dependence and is found
to be $D(z) = D\{1- (\xi_0 |k_\rho\pm k_{\rho'}|)^{-1}\,
[1-e^{-2z|k_\rho\pm k_{\rho'}|}]\}$ with the $\pm$ sign
corresponding to ${\cal Y}^{\rm {D}}_0$ and ${\cal Y}^{\rm
{C}}_0$\,, respectively. Therefore, we justify microscopically the
crucial conjecture--position-dependent diffusion coefficient--made
in Ref.~[\onlinecite{Lagendijk00}] at the one-loop level.
Nevertheless the difference should be stressed that at the medium
boundary $z=0$ the bare diffusion constant is protected against
one-loop WL. Such an important property persists up to higher order
loop corrections enforcing
\begin{equation}
D(z=0)=D \,.
\label{Dboundary}
\end{equation}

In the bulk: $z\gg |k_\rho\pm k_{\rho}'|^{-1} $ the diffusion
coefficients become homogeneous: $D(z) \rightarrow D[1- (\xi_0
|k_\rho\pm k_{\rho}'|)^{-1}]$ but strongly depend on their infrared
cutoffs. By contrast, since Eq.~(\ref{albedo1}) suggests that all
the interfering optical paths contributing to the line shape
reside in a boundary layer of size $\sim |q_\perp|^{-1} \, (\gg
l)$\,, for $0<2\pi/a - |q_\perp|\ll \pi/a$
the local diffusion coefficient of ${\cal Y}^{\rm {C}}_0$ is
simplified as $D(z)\approx D(1-2z/\xi_0)$\,, and well approximated
by $D(z)\approx D/(1+2z/\xi_0)$ as the leading order $l/\xi_0$
correction concerned. Replacing $D$ in Eq.~(\ref{bareproagator})
with the latter we obtain:
\begin{eqnarray}
{\cal Y}^{\rm {C}}_0 = \frac{1}{D\xi_0 |q_\perp|^2} \left[ f_- (x_>)
f_+ (x_<)  - C\, f_- (x_>) f_- (x_<)\right] \,, \label{Ysolution}
\end{eqnarray}
where $x_{>(<)} = \xi_0|q_\perp| [1+\xi_0^{-1}\, {\rm max} \, ({\rm
min}) \, \{z\,,z'\}]$\,, $C=f_+(\xi_0 |q_\perp|)/f_-(\xi_0
|q_\perp|)$ and $f_+(x) = xI_1(x)$ and $f_-(x) = xK_1(x)$ with
$I_1(x)\,, K_1(x)$ the modified Bessel functions.
Eq.~(\ref{Ysolution}) is then inserted into Eq.~(\ref{albedo1}). For
$\xi_0|q_\perp| \gg 1$ one may use the asymptotic expressions of the
Bessel functions and eventually find the line shape given by the
second line of Eq.~(\ref{result})
\cite{noteWL}.
At $|q_\perp|
\sim \pi/a$
interfering optical paths significantly penetrate into the bulk and
suffer from stronger WL, resulting in a larger enhancement factor.

For (ii) higher order expansion of Eq.~(\ref{albedo1}) shows that
the line shape, indeed, is contributed by interfering optical paths
forming the loop-wise structure classified into: diffuson-cooperon
coupling where two paths may trace some loops in the same direction
(e.g, Fig.~\ref{fig3} (a)), and cooperon-cooperon coupling where two
paths trace all the loops in the opposite direction (e.g,
Fig.~\ref{fig3} (b)). As shown below it is the latter leading to SL
in the bulk at $|q_\perp| \sim \xi_0^{-1}$\,, and the line shape at
$|q_\perp| \lesssim \xi_0^{-1}$ is mainly responsible for by
(radiative) SL states with lower symmetric $T$\,.

For (iii) to calculate the lineshape namely ${\cal Y}^{\rm {C}}_0
(z,z';k_\rho,k_\rho')$ exactly is a very hard task. Due to the
broken translational symmetry the technique of Ref.~\cite{Efetov97}
breaks down and the solution there is not applicable. Instead, our
method combines the microscopic yet advanced mathematical theory
namely performing super-Fourier analysis \cite{Zirnbauer91} for
Eqs.~(\ref{albedo1}) and (\ref{DC}) (This is far beyond the scope of
this letter and the details are to be reported elsewhere
\cite{Tianunpub}.) and (phenomenological) hydrodynamic methods
\cite{Vollhardt80}. Observing that $\tilde{\alpha}(\theta) \approx
l^2 \sum_{N} \sum_{k_\rho,k_\rho'}\!\!'^N \,{\cal Y}^{\rm {C}}_N
(l,l;k_\rho,k_\rho')$ we need to consider only the returning
probability-like propagator ${\cal Y}^{\rm {C}}_0
(l,l;k_\rho,k_\rho')$ in (iii).

Applying the heat kernel method \cite{Zirnbauer91} to Eq.~(\ref{DC})
we succeed to calculate ${\cal Y}_0^{\rm C}(z,z;0,0)$ exactly
provided that $T$ is lowered down to the ${\rm {\bf
GMat}}(3,2|\Lambda)$ symmetry \cite{Tianunpub}. The most important
feature of the solution is the exponential divergence
\cite{Tianunpub}: (All the numerical factors are unimportant and not
given here.)
\begin{equation}
\ln {\cal Y}_0^{\rm C}(z,z;0,0) \sim z/\xi_0 + o(e^{-z/\xi_0})
\label{divergence}
\end{equation}
for $z\gtrsim \xi_0$\,. On the other hand, because it was shown that
bulk SL states display hydrodynamic behavior \cite{Vollhardt80} we
expect in the presence of vacuum-medium interface the exponential
divergence Eq.~(\ref{divergence}) to be reflected at the same
macroscopic level. To achieve this we notice that the boundary
leakage introduces the level broadening $\Gamma(z)$ scaling as $-\ln
\Gamma(z)\sim z/\xi_0$ \cite{Chernyak92}. On the physical ground the
contribution of local currents to the restoring force--leading to
SL--exponentially decays in time, modifying the Vollhardt-W{\"o}lfle
model \cite{Vollhardt80} to be
\begin{eqnarray}
\partial_t \, {\cal Y}^{\rm {C}}_0 + \nabla \cdot
{\bf j} &  =  & \frac{1}{\pi\nu}\, \delta(z-z') \,, \label{hydrodynamics} \\
\partial_t \, {\bf j} +\frac{D}{l}\, \nabla {\cal Y}^{\rm {C}}_0 & = &
-\frac{{\bf j}}{l}-\frac{D}{\xi \, l} \int_0^t dt' \, e^{-\Gamma (z)
(t-t')}\, {\bf j}(t') \nonumber
\end{eqnarray}
with ${\cal Y}^{\rm {C}}_0|_{z=0}=0$ and the covariant derivative
$\nabla \equiv (\partial_z\,, \partial_\rho + i q_\perp)$\,, where
the overall coefficient of the restoring force term is fixed by
Eq.~(\ref{Dboundary}). Solving Eq.~(\ref{hydrodynamics}) indeed
confirms Eq.~(\ref{divergence}) ($q_\perp=0$ and the localization
length $\xi=\xi_0$).

The presence of nonvanishing $k_\rho\approx -k_\rho'$ alters the
microscopic symmetry and therefore the localization class. Indeed,
(for $|q_\perp| \lesssim \xi_0^{-1}$) one may follow
Ref.~\cite{Altshuler93}, average Eq.~(\ref{DC}) over $|k_\rho -
k_{\rho}'|$
and subsequently obtain an effective action: $F[Q]=\frac{\pi a\nu
D}{8}\int_0^\infty dz \, {\rm str} \{(\partial_z Q)^2 + [iq_\perp
\Lambda/2 , Q]^2 \}$\,. The $T$-field symmetry is lowered down to
$U(1,1/2)/U(1/1)\times U(1/1)$ giving $\xi=2\xi_0$ \cite{Efetov97}.

On the other hand, the common belief of one-parameter scaling
hypothesis (e.g. Ref.~\cite{Woefle02}) implies that the microscopic
symmetry enters only through the localization length leaving the
hydrodynamic model unaffected. To find ${\cal Y}^{\rm {C}}_0
(z,z';k_\rho,k_\rho')$ we insert the $z$-dependent level broadening:
$\Gamma(z) = D/(2\xi_0)^2\, e^{-\frac{z}{2\xi_0}}$ \cite{
Chernyak92} and $\xi=2\xi_0$ into Eq.~(\ref{hydrodynamics}). For
$t\gg \xi_0^2/D$ the steady distribution is approached solving:
$-D\{\partial_z e^{-\frac{z}{2\xi_0}}
\partial_z - |q_\perp|^2 e^{-\frac{z}{2\xi_0}}\}\, {\cal Y}^{\rm {C}}_0
= (\pi\nu)^{-1}\, \delta (z-z')$\,, which coincides with the
diffusive model for the single channel SL \cite{Lagendijk00}.
Substituting the solution  and the bare propagator,
Eq.~(\ref{bareproagator}) into Eq.~(\ref{albedo1}) one may find
\begin{eqnarray}
\! \tilde{\alpha}(\theta)
&  =  &
\left(\!1 \! - \! \frac{a|q_\perp|}{2\pi}\!\right) \!\!\left[1\! +
\! \frac{3\! - \!\sqrt{1\! + \!(4\xi_0 q_\perp)^2}}{2\xi_0/l}\right]
\! + \!
 \frac{a|q_\perp|}{2\pi}
I_0 (\theta) \nonumber\\
& \rightarrow & \left(1+l/\xi_0\right)-a/(2\pi\xi_0)\, l|q_\perp|
\,, \quad |q_\perp| \ll \xi_0^{-1} \,. \label{albedolocalization}
\end{eqnarray}
The first line holds for $|q_\perp|\lesssim \xi_0^{-1}$ and, apart
from the factor: $[1-a|q_\perp|/(2\pi)]$ accounting for the weight
of quasi-$1$D motion, the first term resembles the rounded line
shape
of fully disordered media below localization transition
\cite{Lagendijk00}. Remarkably, the second line suggests that the
line shape displays a blunt triangular peak in a very narrow region:
$\quad |q_\perp| \ll \xi_0^{-1}$ (inset of Fig.~\ref{fig2}).




{\it Conclusions.}  Analytical studies of the coherent
backscattering line shape have been presented for periodic thick
disordered medium films which arrest
both extended and quasi-$1$D localization states.
The result is expected to be qualitatively correct for
$\lambda\lesssim l \lesssim a \, (\lesssim l e^{l/\lambda})$ which,
together with the realization of the perfect periodicity along one
direction, may be well within the reach of up-to-date experimental
conditions \cite{Xu06}. However, to study realistic media the
present theory still needs to be extended so that the large size
(namely the film thickness much larger than the lattice constant)
effects and the parity (mirror) symmetry are taken into account,
which is left for future work.

I am grateful to S. Hikami for useful conversations, especially A.
Altland, L. Zhou and M. R. Zirnbauer for important discussions. Work
supported by Transregio SFB 12 of the Deutsche
Forschungsgemeinschaft.


\begin{thebibliography}{}
\bibitem{notecondition}
Medium gain/loss and internal reflections are ignored.
\bibitem{notefactor} The overall factor: $l^3/(2D)\,
(\sum_{k_\rho,k_\rho'}^{'N} \, 1+ \sum_{k_\rho,k_\rho'}^{'N+1} \,
1)=l^3/(2D)\,(\sum_{k_\rho} \!\! 1)$ is dropped out hereafter.
\end{thebibliography}
\end{document}